\documentclass[runningheads]{llncs}
\usepackage{booktabs}
\usepackage{array}
\usepackage{xurl}

\usepackage[T1]{fontenc}
\usepackage{graphicx}
\usepackage[]{subfig}
\usepackage{xcolor}
\usepackage{comment}
\usepackage[misc,geometry]{ifsym}

\usepackage{hyperref}
\usepackage{ragged2e}

\newcommand{\accessed}{(Accessed: 2025-06-10)}

\begin{document}
\title{Transforming Football Data into Object-centric Event Logs with Spatial Context Information}
\titlerunning{Transforming Football Data into OCEL with Spatial Context Information}
\author{Vito Chan\inst{1,2}$^{\textrm{(\Letter)}}$
\and
Lennart Ebert\inst{1}
\and
Paul-Julius Hillmann\inst{1}
\and
Christoffer Rubensson\inst{1,2}
\and
Stephan A. Fahrenkrog-Petersen\inst{1,2}
\and
Jan Mendling\inst{1,2,3}
}
\authorrunning{Chan et al.}
\institute{Humboldt-Universität zu Berlin, Berlin, Germany
\and
Weizenbaum Institute, Berlin, Germany
\and
Wirtschaftsuniversität Wien, Vienna, Austria
\\
\email{firstname.lastname@hu-berlin.de}
}
\maketitle              %
\begin{abstract}
Object-centric event logs expand the conventional single-case notion event log by considering multiple objects, allowing for the analysis of more complex and realistic process behavior. However, the number of real-world object-centric event logs remains limited, and further studies are needed to test their usefulness. The increasing availability of data from team sports can facilitate object-centric process mining, leveraging both real-world data and suitable use cases. In this paper, we present a framework for transforming football (soccer) data into an object-centric event log, further enhanced with a spatial dimension. We demonstrate the effectiveness of our framework by generating object-centric event logs based on real-world football data and discuss the results for varying process representations.
With our paper, we provide the first example for object-centric event logs in football analytics. Future work should consider variant analysis and filtering techniques to better handle variability.

\end{abstract}

\section{Introduction}
\label{sec:introduction}

Object-centric process mining~\cite{van2019object} is focussed on analyzing complex systems with multiple objects that interact with each other through their behavior. So far, real-world data of this type is scarce due to confidentiality and privacy concerns~\cite{ElkoumyFSKMVRW22}.
In the past, alternative data sources have been proposed to address these issues. For example, Liss et al.~\cite{liss2024framework} utilized data from the strategy game \textit{Age of Empires 2} to generate object-centric event logs. However, there is still a lack of object-centric event data and studies that evaluate its usefulness.

Team sports offer an excellent setting for object-centric process mining.
Not only because of its sequential nature~\cite{gudmundsson2017teamsportsslr}, which involves multiple objects to consider, such as teams, players, and game commodities, but also due to the relatively high number of publicly available football (soccer) data. Data-driven methods for analyzing sports data are gaining popularity (cf., \cite{gudmundsson2017teamsportsslr,wakelam2022footballattrslr}) in fields such as visual computing and visual analytics~\cite{gudmundsson2017teamsportsslr,naik2022sportscomputervisionslr}. In process mining, work that considered football data shows the benefits of applying process-centric techniques to analyze football tactics~\cite{caparros.2023,krockel2020footballPM}. Still, these works only consider conventional event logs with a single case notion, where the ball possession of a team was modeled as a specific case. 
Limiting analysis to on-ball actions ignores other behaviors that could be important to understand game performance, such as those of an idle player. 
In addition, the object-centric notion could better reflect the complexity of football games, while also addressing issues such as convergence and divergence (cf.,~\cite{van2019object}).
However, work that considers object-centric event logs is lacking. 

In this paper, we present a framework to transform football data into object-centric event logs. By considering multiple objects, such as teams, players, and the ball, we enable the modeling of more comprehensive behaviors in football games using process mining. We also consider spatial information, as this can further support analysis, such as investigating behaviors at different positions on a football field. We evaluate the framework by generating an object-centric event log from a real-world dataset and qualitatively comparing how changing the number of objects impacts the analysis at both the process and instance levels of the data. Future work could consider combining object-centric process mining with variant analysis and filtering techniques to handle the high complexity and variability of such logs.

The remainder of this paper is structured as follows. \autoref{sec:background} discusses the theoretical foundations. \autoref{sec:method} presents our framework. 
\autoref{sec:results} validates our framework with real-world data. \autoref{sec:discussion} discusses the findings, and \autoref{sec:conclusion} concludes this paper.

\section{Background}
\label{sec:background}

In this section, we first describe object-centric process mining in \autoref{sec:back_ocel}. Then, in \autoref{sec:back_football}, we discuss previous work on football analytics and process mining, as well as provide an overview of football data. We conclude with a problem statement in \autoref{sec:problemstate}.

\subsection{From Case-centric to Object-centric Process mining}
\label{sec:back_ocel}

Process mining techniques derive process-centric insights from event data~\cite{aalst2022deg360}. Event data are records of events of a process, partially ordered in an event log.
Each event is associated with at least an \emph{event id}, an \emph{event type} (activity), a \emph{timestamp}, and a \emph{case id}. Typically, other attributes specific to the process are included, such as information about where an activity was executed and by whom. 
A conventional event log refers only to a single case notion\footnote{We will use the terms \emph{case notion} and \emph{object type} interchangeably.}, i.e., all cases are of the same type to which all events apply. For example, in an order-to-cash process, all cases could be referred to orders. A single order is then a sequence of events that describes the activities executed from order placement to payment and delivery.

Real-world processes are complex, and events can be associated with multiple case notions. For example, the events of an order-to-cash process could be associated with one or more object types, such as an order, item, package, and/or route~\cite{van2019object}. Enforcing events to be associated with only one of these notions can lead to missing events (deficiency), unintentional duplications of events (convergence), or unclear causal dependencies between events related to multiple case notions (divergence)~\cite[p. 20]{aalst2022deg360}. Hence, instead of referring to a single case notion, as in conventional event logs, an object-centric event log allows events to be associated with multiple case notions. In this way, object-centric process mining can provide a more comprehensive view of the process while mitigating some of the issues associated with traditional process mining based on conventional event logs. For instance, returning to the earlier example, instead of treating each order as completely separate, object-centric process mining can capture their interrelations and connections to other objects and object types --- such as a shipping event and a truck fulfilling a delivery on a certain route, which could be related to multiple orders.

\subsection{Football Analysis}
\label{sec:back_football}

Football analysis research is an emerging field that is receiving increasing attention, with multiple studies already summarizing the current state of knowledge. Sarmento et al.~\cite{sarmento2018performance,sarmento2014match} conducted two systematic reviews to identify the most common topics in match analysis research up to and including 2016. In~\cite{sarmento2014match}, they grouped work based on the type of analysis: \emph{descriptive}, \emph{comparative}, and \emph{predictive}. They found that most studies investigate performance by describing player behavior while ignoring situational or socio-contextual characteristics~\cite[p. 9]{sarmento2014match}. In~\cite{sarmento2018performance}, they investigated studies in the analytical topics \emph{set plays}, \emph{activity profiles}, and \emph{group behavior}. They found an increasing use of spatiotemporal data to explore behavioral patterns in groups~\cite[p. 833]{sarmento2018performance}. In more recent surveys, Goes et al.~\cite{goes2021bigdatasoccerslr} summarized studies on tactical behaviors that employed position-tracking data. They identified methodological differences between sports science and computer science, which could present opportunities for future work through interdisciplinary work. Moreover, Wakelam et al.~\cite{wakelam2022footballattrslr} focused specifically on studies that analyze player attributes. They found that the majority of the studies applied statistical techniques to analyze player traits, rather than, for example, machine learning techniques~\cite[p. 62]{wakelam2022footballattrslr}.

In recent years, there has also been an increase in data-driven sports analysis (cf., \cite{gudmundsson2017teamsportsslr,wakelam2022footballattrslr}) — utilizing either event data or trajectory data~\cite{gudmundsson2017teamsportsslr}. This opens up opportunities for applying data mining techniques such as those in process mining. In the following subsections, we provide an overview of football logs and then discuss works that have applied process mining techniques to such data.

\subsubsection{Football Logs}

\begin{table}[ht]
   \caption{An overview of football datasets.}
    \label{tab:soccer_logs_overview}
    \centering
    \scriptsize
\resizebox{\textwidth}{!}{%
\begin{tabular}
{@{}>{\raggedright\arraybackslash}p{0.3\linewidth}>
{\raggedright\arraybackslash}p{0.15\linewidth}>{\raggedright\arraybackslash}p{0.1\linewidth}>{\raggedright\arraybackslash}p{0.15\linewidth}>{\raggedright\arraybackslash}p{0.15\linewidth}>{\raggedright\arraybackslash}p{0.15\linewidth}@{}}
 
\toprule
Event log                             & \# of Games           & Ball actions & Non-ball actions (fouls, cards etc.) & Player movements & Availability \\ \midrule
StatsBomb open data & \textgreater{}1000    & Yes & Yes & No & Public\textsuperscript{\ref{fn:statsbomb}}
 \\
OPTA Sports \textit{(England-Iceland Euro 2016)} & 1                     & Yes & Yes & No & Private \\
Wyscout dataset                       & \textgreater{}1000    & Yes & Yes & No & Public \textsuperscript{\ref{fn:wyscout}}\\
DFL dataset                           & 7                     & Yes & Yes & Yes (25 Hz) & Public \textsuperscript{\ref{fn:dfl}}\\
Metrica Sports sample data            & 3                     & Yes & Yes & Yes (25 Hz) & Public  \textsuperscript{\ref{fn:metrica}}\\
\bottomrule
\end{tabular}
}
\end{table}

\addtocounter{footnote}{+1} %
\footnotetext[\value{footnote}]{\url{https://github.com/statsbomb/open-data/}\label{fn:statsbomb} \accessed.}
\addtocounter{footnote}{+1} %
\footnotetext[\value{footnote}]{\url{https://figshare.com/collections/Soccer_match_event_dataset/4415000/2}\label{fn:wyscout} \accessed.}
\addtocounter{footnote}{+1} %
\footnotetext[\value{footnote}]
{\url{https://springernature.figshare.com/articles/dataset/An_integrated_dataset_of_spatiotemporal_and_event_data_in_elite_soccer/28196177}\label{fn:dfl} \accessed.}
\addtocounter{footnote}{+1} %

\footnotetext[\value{footnote}]{\url{https://github.com/metrica-sports/sample-data}\label{fn:metrica} \accessed.}

\autoref{tab:soccer_logs_overview} provides an overview of football datasets.
Multiple football event logs have been used by previous research, some of which are publicly available
(cf., \autoref{tab:soccer_logs_overview}). 
The most comprehensive datasets are provided by professional data providers Statsbomb and Wyscout, each including more than 1,000 games. These event logs consist of discrete ball actions (e.g., passes, shots, and free kicks) as well as events not directly related to the ball (e.g., fouls, and cards). 

Furthermore, two recently released datasets, DFL and Metrica Sports, also include player movement data. By automated visual player tracking, these datasets report all player positions at a rate of 25 Hz (i.e., 25 recorded player positions per second).
The Metrica Sports dataset encodes the positions of all players as $(x, y)$ coordinates in a vertically flipped Cartesian coordinate system, with $(0, 0)$ indicating the origin in the top-left corner, whereas $(1, 1)$ is located in the bottom-right corner.

\subsubsection{Football analysis in Process mining}

Only a couple of works have considered football logs in process mining. Kröckel and Bodendorf~\cite{krockel2020footballPM} were the first to demonstrate how process mining techniques, in combination with visual analytics, can be applied to football logs to support better decision-making. By applying multiple process mining techniques such as discovery and social network analysis on the OPTA Sports dataset (see \autoref{tab:soccer_logs_overview}), they found that process mining methods can provide analysts with user-friendly options to derive a wide range of insights about game success and game tactics from different perspectives. 
Furthermore, Caparrós~\cite{caparros.2023} proposed a process discovery methodology to analyze team behavior. The methodology employed a \emph{purpose-oriented} trace filtering technique that reduces variability, thereby making the analysis more tailored for such data while handling some of its complexity. They demonstrated their methodology using the StatsBomb dataset (see \autoref{tab:soccer_logs_overview}).

\subsection{Problem Statement}
\label{sec:problemstate}

Current work on process mining for football analysis is limited in that it focuses on players directly involved in on-ball actions and does not account for the impact of other players on game performance. However, all players, including those who are idle, may also play a significant role in the game's success. Moreover, the traditional event log for football data also suffers from the problems of divergence and convergence when attempting to capture the complexity of the game under a single case notion. Hence, this study aims to address these issues by applying the notion of object-centric process mining.

\section{Object-Centric Framework for Football Event Data}
\label{sec:method}

\begin{figure}
    \centering
    \includegraphics[width=0.75\linewidth]{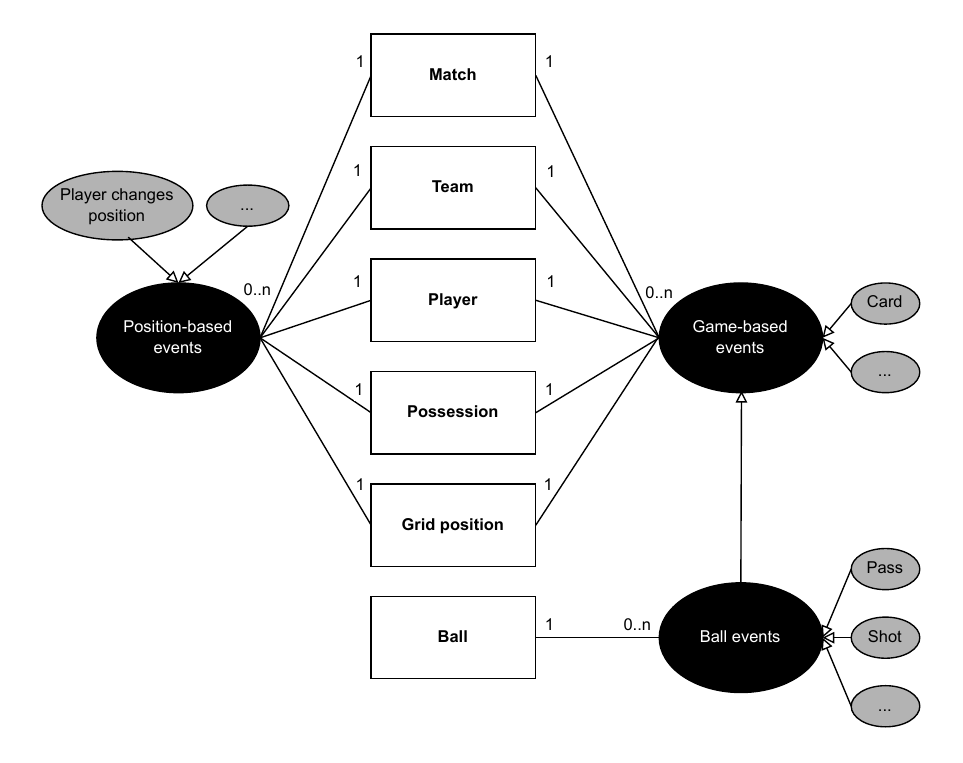}
    \caption{Object-Centric Framework for Football Event Data}
    \label{fig:er-diagram}
\end{figure}

In this section, we propose a novel object-centric framework for representing football match data. 
This framework includes football-specific objects, events and object-event relationships that enable multi-variate process mining analysis.

\autoref{fig:er-diagram} depicts the proposed object-centric framework. It defines three event classes: \textit{game-based events}, \textit{ball events}, and \textit{position-based events} (shown as black ellipses).
Light gray ellipses illustrate concrete examples of activities that belong to a corresponding event class.
The six object types are: \textit{match}, \textit{team}, \textit{possession}, \textit{player}, \textit{grid position} and \textit{ball} (shown as squares).
\textit{Ball events} can inherit all object types from game-based events, illustrated by an arrow.
The cardinalities indicate that each event must involve all object types to which it is connected.
In the following, we discuss the individual event classes and object types.

\subsection{Event Classes}

The framework defines three event classes that each contain related event types.
Example event types are given in \autoref{fig:er-diagram}. The actual event types depend on the source data.
The first event class is \textit{game-based events}. These include referee events, such as yellow cards, as well as tackles. 
The second event class is \textit{ball events}, which are specialized game-based events that involve a ball. Passes, shots, and free kicks are examples of ball actions in football.
The third event class is \textit{position-based events}. These include all types of events related to player positions.
Our concrete implementation of the framework includes the ``Player changes position'' activity as part of position-based events.
Other implementations may include all player positions at all times.

\subsection{Object Types}
In the following, we present each object type in detail, describe how they relate to the different event classes, and provide example analytical questions for each object.
Although each question is associated with a single object, they can be adapted or combined to investigate intra- or inter-related object behaviors, such as interactions between players or between player and ball.

\subsubsection{Match}
Matches represent separate games. An example game involves two teams competing against each other at a specific date, such as VfL Wolfsburg playing against Bayern Munich on August 25, 2024. The match object provides an overview of the process of an entire game, hence can answer questions such as:
\begin{enumerate}
   \item How many goal shots occurred in a specific match?
   \item How many matches did a particular player play in?
   \item What tactical movement patterns show after a goal is scored against a team?
\end{enumerate}

\subsubsection{Team}
Teams compete with each other, either during a specific match, in a league or in a tournament. Examples for teams are Bayern Munich, the German national team or a university's football club. With teams as objects, we can answer team-level questions such as:
\begin{enumerate}
    \item How did team tactics evolve across multiple games?
    \item What was the interaction pattern between two teams during a match (e.g., how did one team typically react to another team's movement)?
    \item How much did a team's players move during a match?
\end{enumerate}

\subsubsection{Player}
Players represent the individual actors who participate in various events, including passes (ball events), card assignments (game-based events) or positional movements (position-based events).
The \emph{player} object type enables analysis of dependencies between players and player behavior:
\begin{enumerate}
    \item Which player did one specific other player typically pass to?
    \item How much did a player move during a match?
    \item What was the running path of a player before intercepting the possession of the other team?
\end{enumerate}

\subsubsection{Possession}
A possession starts when a team gains control over the ball and ends when the opposing team gains control. Thereafter, the next possession starts.
During a possession, one team is attacking, while the other is defending the ball.
E.g., at the very start of a football game, the first possession begins with the kickoff. If the team that kicked off loses the ball to the other team, the first possession ends, and the second possession begins.
Possessions can be used to identify offensive and defensive plays and allow for analysis such as:
\begin{enumerate}
    \item What activity variants led to possessions that resulted in a goal being scored?
    \item How did the average running distance per possession vary between teams?
    \item What was the average number of passes per possession?
\end{enumerate}

\subsubsection{Grid position}
\begin{figure}[ht]
    \centering
    \includegraphics[width=0.4\linewidth]{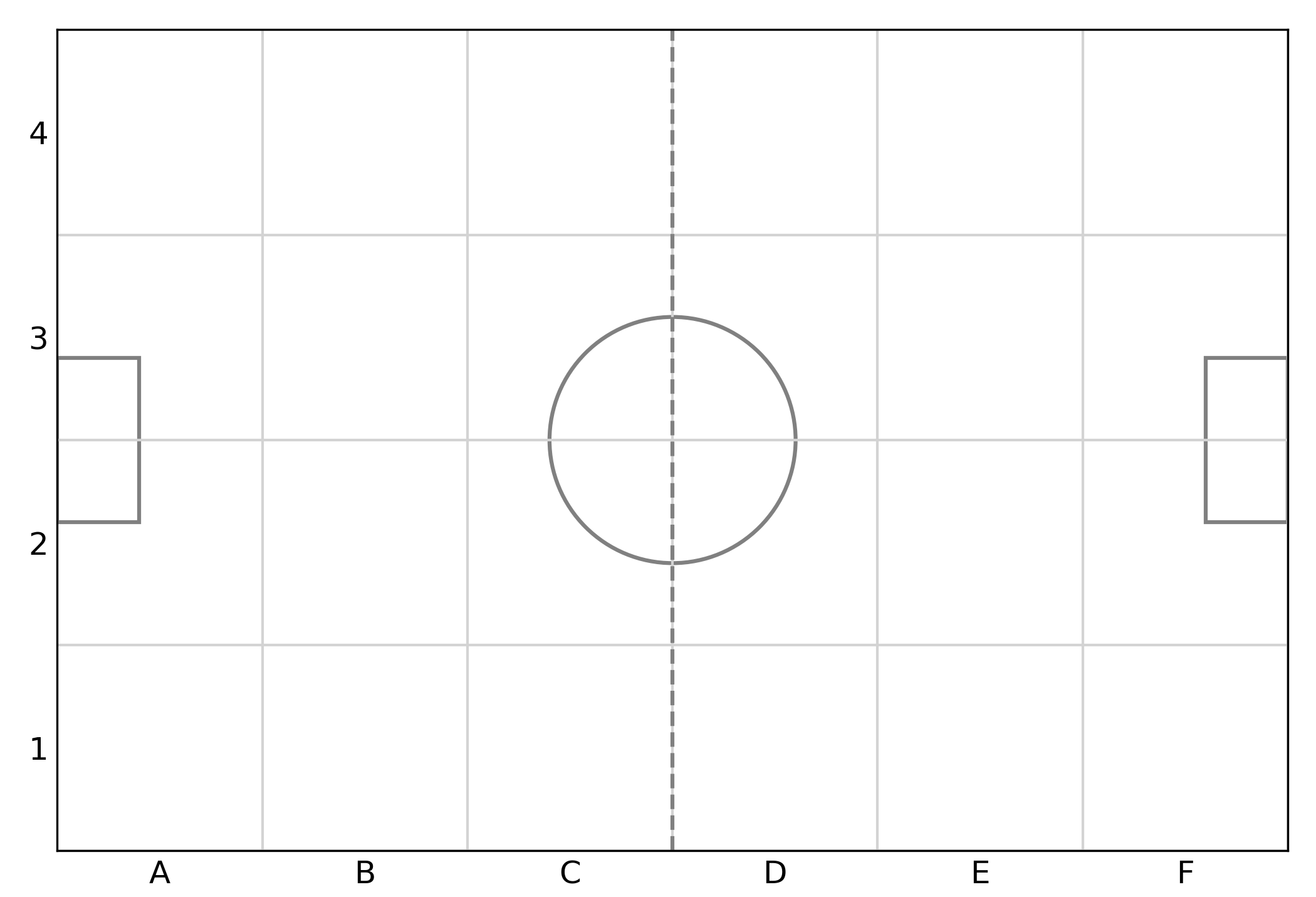}
    \caption{Extracted grid positions (6 by 4)}
    \label{fig:grid_pos}
\end{figure}

The football field can be divided into a grid with individual grid positions, simplifying spatial analysis.
For our purposes, we consider a six-field-wide and four-field-long grid, as this greatly reduces the spatial complexity while still maintaining basic positional information. \autoref{fig:grid_pos} illustrates such a grid. The six-by-four grid results in 24 grid positions numbered A1 (bottom left) to F4 (top right).
Grid positions are recorded for all events, denoting the location where each event occurred.
Modeling grid positions as objects instead of attributes allows for queries such as:
\begin{enumerate}
    \item How many goal shots occurred from a specific grid position?
    \item Which is the player that spends the most time on a specific grid position?
    \item How long are players in specific grid positions?
\end{enumerate}

\subsubsection{Ball}
The ball is the central object that moves across the field during a football match.
It is contained in all ball events (e.g., passes).
By modeling the ball as a separate object type, we can answer questions such as:
\begin{enumerate}
    \item What activities happened to the ball before a goal was scored?
    \item Which players played the ball most often?
    \item How frequent are ball passes vs. shots?
\end{enumerate}

\section{Evaluation}
\label{sec:results}
In this section, we validate our framework. Specifically, we first generate an object-centric event log using real-world data. Then, we qualitatively analyze the data by comparing a single-object type version with a multi-object type version of the log from a process perspective using a directly-follows graph, as well as from an instance perspective, with instances represented as graphs on a spatial map. \autoref{sec:implementation} describes the applied dataset and log generation process, whereas Sections \ref{sec:processview} and \ref{sec:instanceview} discuss evaluation results.

\subsection{Event Log Generation}
\label{sec:implementation}

We applied the data of two football matches from the Metrica Sports sample data (see Table~\ref{tab:soccer_logs_overview}) to generate an object-centric event log using our framework in the following steps.
First, we recognized meaningful player movement events by reprojecting the player movement traces onto the grid coordinate system and identifying movements across grids. We used Pandas\footnote{\url{https://pandas.pydata.org} \accessed.} to process and transform the raw data. Second, we performed event engineering by decomposing aggregated events to improve interpretability and clarify the association between specific objects and events. We then merged the movement and game-based event logs based on temporal order and propagated the contextual information to enrich the movement events. Finally, we further enriched the event log with derived attributes such as travel distance, durations, and the current score. The final processed \emph{DataFrame} was converted into an object-centric event log using PM4Py\footnote{\url{https://processintelligence.solutions} \accessed.}. 
The resulting log comprises 37,358 events, 813 objects, and 747 possessions, capturing both game-based events and position-based events. 
The source code to generate the event logs and perform subsequent visualization is publicly available on GitHub.\footnote{\url{https://github.com/VitoChan01/Soccer}.}

\begin{figure}[!ht]
\centering
    \subfloat[\emph{Ball} object type \label{fig:trad_dfg}]{\includegraphics[width=0.8\linewidth]{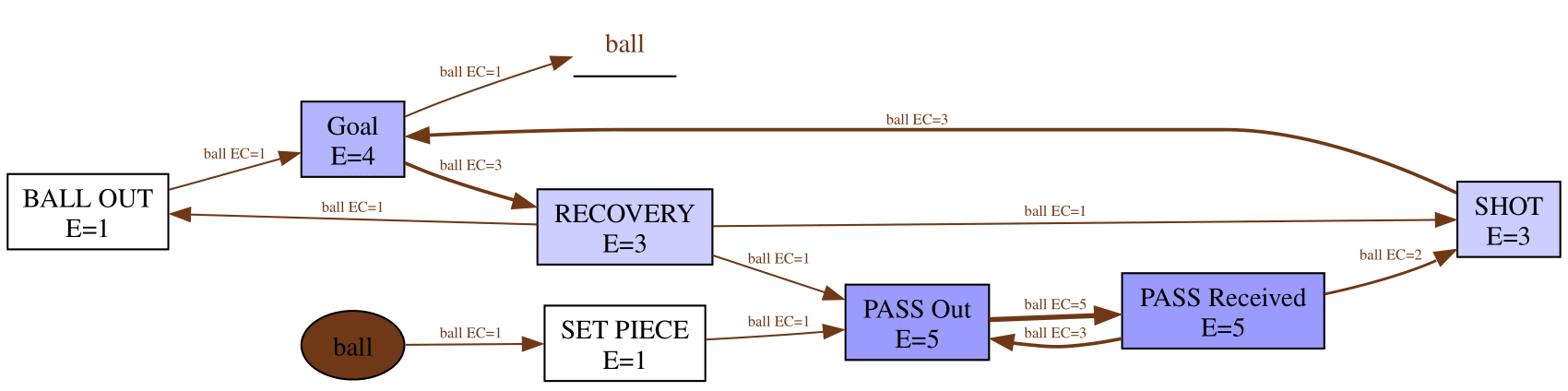}} \\
    \subfloat[\emph{Ball}, \emph{possession} and \emph{player} object types \label{fig:ocdfg}]{\includegraphics[width=\linewidth]{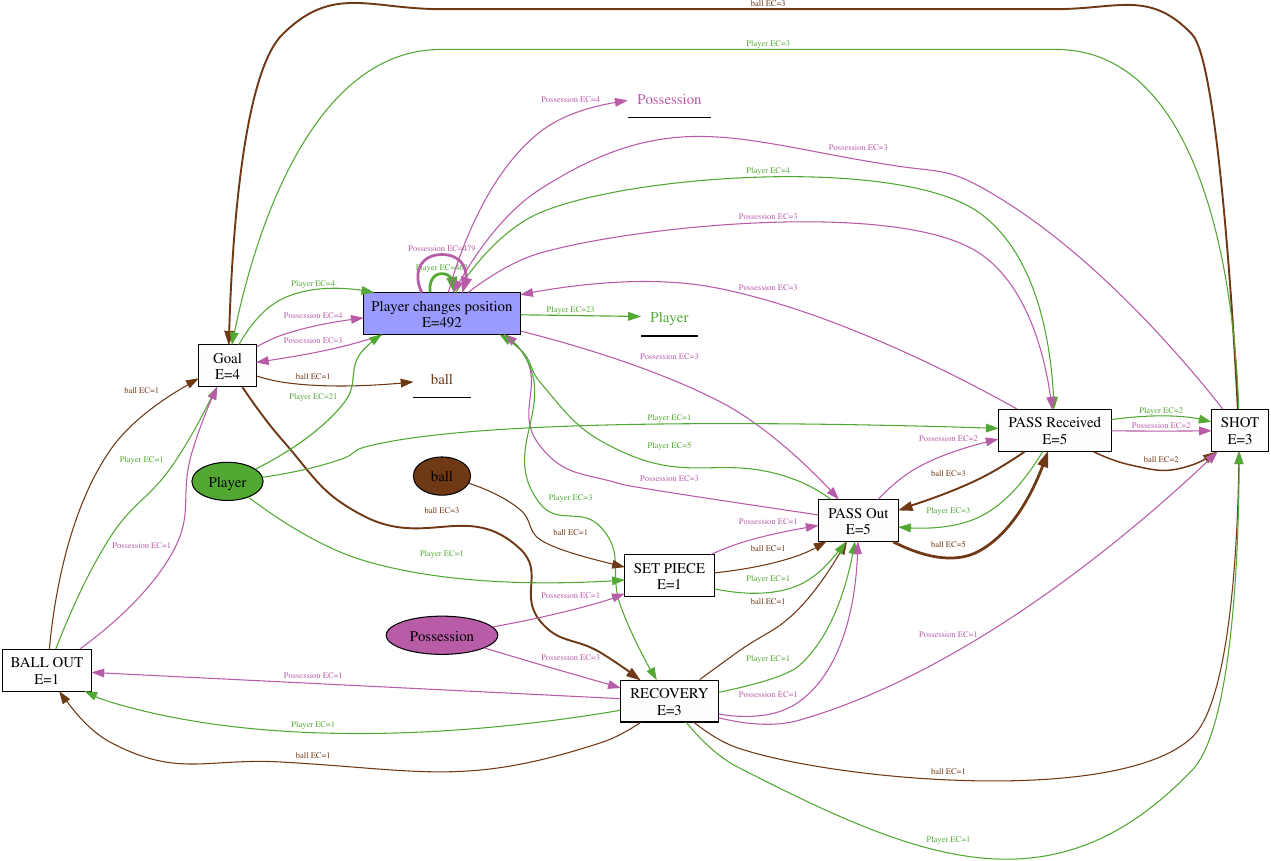}} 
    
    \caption{Discovered directly-follows graphs from the object-centric event log.}
    \label{fig:spatialView_comparison}
\end{figure}

\subsection{Perspective 1: Process Model}
\label{sec:processview}

Figures~\ref{fig:trad_dfg} and \ref{fig:ocdfg} present directly-follows graphs discovered from a single-object type event log and a multi-object type event log, respectively. To reduce their complexity, both graphs are filtered to the four possessions of the home team that led to a goal.
The first graph discovered from the single-object type event log (Figure~\ref{fig:trad_dfg}) depicts the sequences of actions related to the \emph{ball} object type, i.e., all on-ball actions such as passes. The graph shows, for instance, that \textit{set piece} activities (e.g., kick-offs, free-kicks, and throw-ins) are always followed by playing a pass. After one or multiple passes, there is a shot. For the possessions in this scenario, all shots led to a goal.
The data also shows a recovery activity leading to an out ball and, subsequently, a goal. However, this behavior is likely caused by erroneous data. Furthermore, note that as we depict the \emph{ball} object type, there are also goals followed by a recovery, since the ball remains the same throughout the entire game. 

The multi-object type process graph (Figure~\ref{fig:trad_dfg}) additionally shows \emph{player} and \emph{possession} objects. Further object types were omitted to avoid visual complexity. 
This graph provides additional information not visible in the previous graph, such as new directly-follows relations for the different object types, like \emph{player}.
Also, an additional activity, ``Player changes position'', is now visible, as it is associated with these two object types.
This activity shows a high count of self-loops, as many players might move one or multiple positions between other activities.

\subsection{Perspective 2: Process Instances on Spatial Map}
\label{sec:instanceview}

\begin{figure}[ht]
\centering
    \subfloat[\emph{Ball} object \label{fig:spatialView_traditionalLog}]{\includegraphics[width=0.5\linewidth]{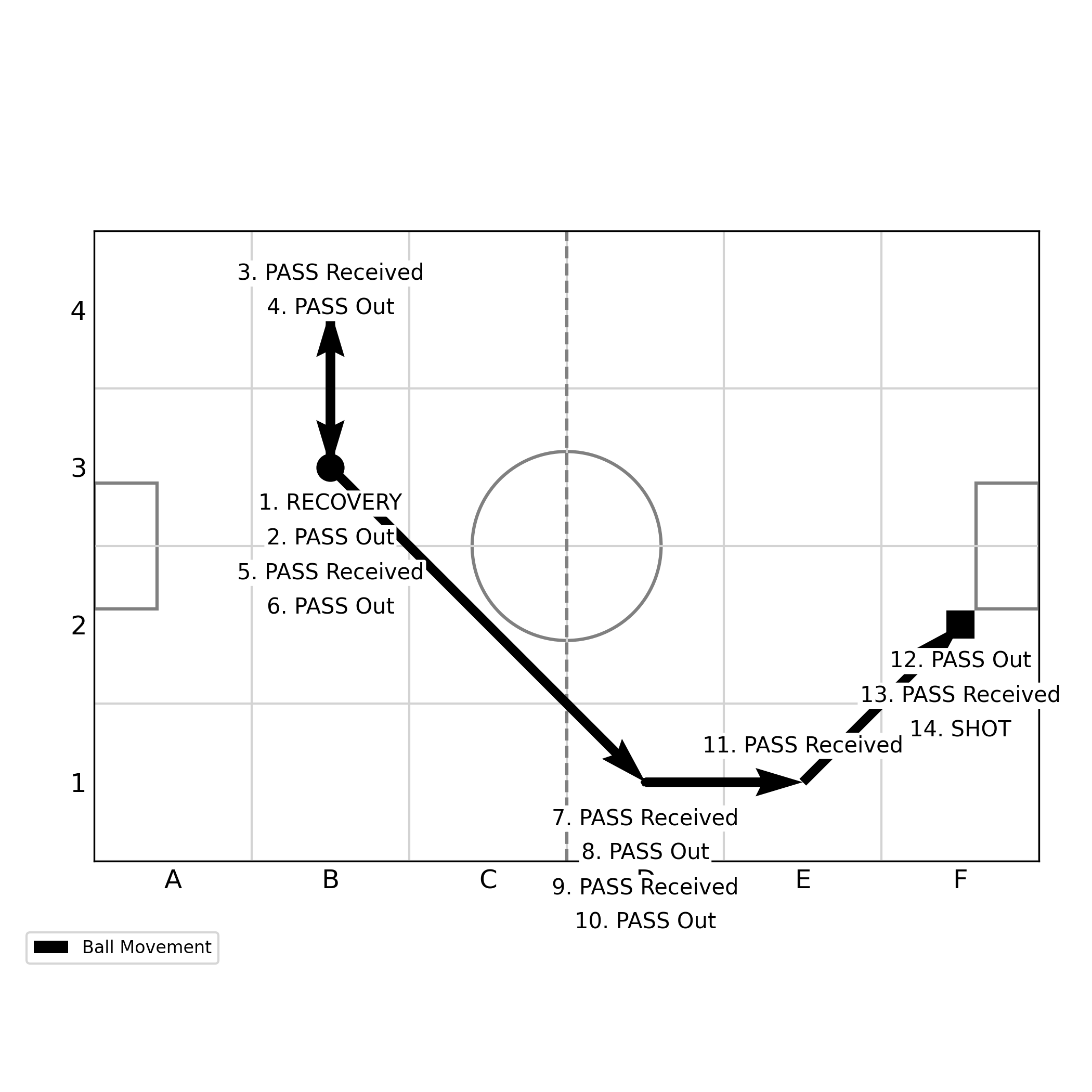}}
    \subfloat[\emph{Ball} and \emph{player} objects \label{fig:spatialView_OCEL}]{\includegraphics[width=0.5\linewidth]{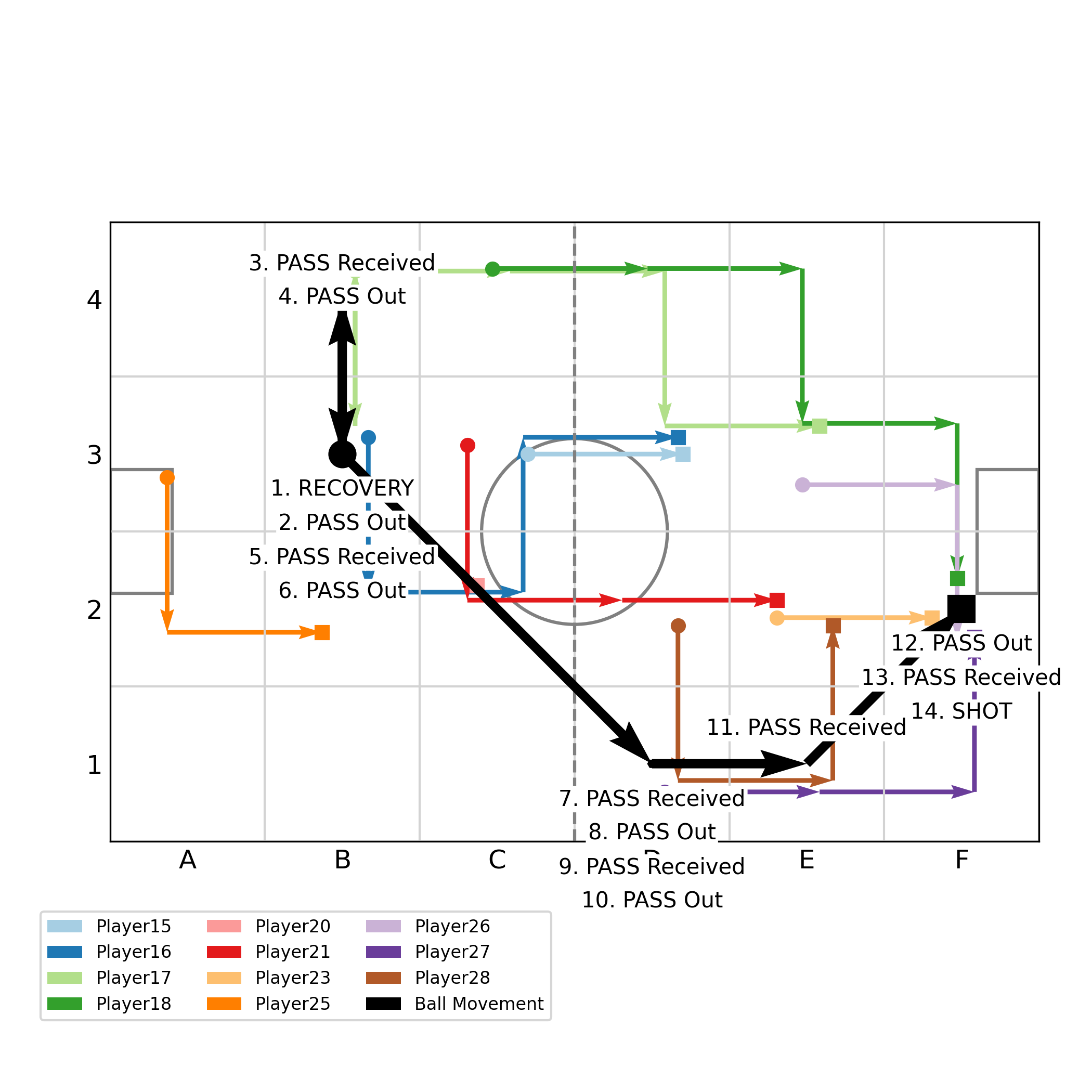}} 
    
    \caption{
    Process instances, visualized as directed graphs on a spatial map.}
    \label{fig:spatialView_comparison}
\end{figure}

Based on the grid notion described in \autoref{sec:method}, we created a custom visualization to display the spatial positions of events on the football field. Specifically, we visualized all events associated with a team for a single process instance of a possession that resulted in a shot on the opponent's goal (possession id: AA156). The first visualization (Figure~\ref{fig:spatialView_traditionalLog}) illustrates the process instance from the perspective of a single object (\emph{ball}). Whereas the second visualization (Figure~\ref{fig:spatialView_OCEL}) is based on multiple objects (\emph{ball} and \emph{player}). 

In Figure~\ref{fig:spatialView_traditionalLog}, depicting a single \emph{ball} object, we can observe where the possession of the ball started, and how the ball moves across the grid cells of the field until the possession ends. Here, the possession begins with a ball recovery in the middle of the field, located in cell $B3$. It is then passed to $B4$, where it is received and passed back to $B3$. Then, the ball is passed out again and received on the right side in the opponent's half of the field. Here, the ball is first passed within the grid cell and then passed out to $E1$ where the pass is received, moving closer to the opponent's goal. The next on-ball action happens in $F2$, meaning that the player who received the ball in $E1$ moved to $F2$ before the ball is passed within that grid cell and then shot toward the opponent's goal.

In Figure~\ref{fig:spatialView_OCEL}, the same on-ball behavior can be observed. 
However, as this includes the addition of \emph{player} objects, the figure also provides information on the movement of individual players, allowing for a more detailed analysis. For instance, the visualization shows that \textit{Player15} moved only from grid cell $C3$ to $D3$ during the entire possession and therefore stayed relatively stable in the center of the field. In contrast, \textit{Player18} moved all the way from grid cell C4 in the left midfield to grid cell $F2$ where the shot on the opponent's goal was taken, highlighting a higher involvement in the offense. In addition to single-player insights, an overall forward movement of all players in the direction of the ball is also visible during the possession.

\section{Discussion}
\label{sec:discussion}

This section discusses the findings from the evaluation and the limitations of our work. By generating an object-centric event log and analyzing it from two different perspectives, we could demonstrate the effectiveness of our framework. Compared to a conventional event log, modeling multiple object types for football data provides a more comprehensive and realistic view of a football game, supporting a better understanding of tactics and game behaviors. The explicit modeling of spatial information further extends the analytical possibilities, as it enriches the control-flow perspective with important contextual cues for examining idle players (cf., \autoref{sec:instanceview}), which are essential for understanding tactical decisions.

We highlight two limitations as motivations for future work. First, football games are, from a process perspective, complex and highly variable. Object-centric models can better capture this complexity; however, techniques for reducing variability and extracting outcome-centric patterns are necessary to enhance understanding. Future work could consider incorporating variant analysis techniques, filtering methods, or interactive visualization. Complementing the spatial perspective with \emph{time} could also improve analysis by examining specific temporal behavior, such as episodes of a game.

Finally, we have only considered football games and a single dataset in this paper. Although our framework is general enough to be applied to other team sports that involve a ball, adaptations may be necessary to facilitate a more specific analysis of different sports types. The effectiveness of object-centric process mining for football analysis and other team sports also needs to be tested using various real-world event datasets in future work.

\section{Conclusion}
\label{sec:conclusion}
In this paper, we introduced the first study concerned with transforming football data into an object-centric event log. The proposed framework considers object types essential for analyzing the performance of football games, including matches, teams, possessions, players, the ball, and grid positions. 
We validated our framework using a real-world football log, combined with a qualitative analysis of different process visualizations of that log with varying amounts of object types, to demonstrate its effectiveness. In this way, we demonstrated that object-centric process mining of football data provides a more complete and detailed analysis of football games. Still, analyzing football games is challenging due to their high variability. Hence, future work should consider reducing variability with variant analysis and filtering techniques.

\begin{credits}
\subsubsection{\ackname} 
The research of the authors was supported by the Einstein Foundation Berlin under grant EPP-2019-524, by the German Federal Ministry of Research, Technology and Space under grant 16DII133, and by the Deutsche Forschungsgemeinschaft under grants 496119880 (VisualMine), 531115272 (ProImpact), and 414984028 (FONDA).

\subsubsection{\discintname}
The authors have no competing interests to declare that are relevant to the content of this article.
\end{credits}

\bibliographystyle{splncs04}
\bibliography{bibliography}

\end{document}